\documentclass[a4paper]{article}
\usepackage{graphicx,psfrag,amsmath,latexsym}
\renewcommand{\theequation}{\arabic{section}.\arabic{equation}}
\renewcommand{\d}{\partial}
\newcommand{\Int}{\!\int\!}

\newcommand{\op}[1]{\textbf{#1}}
\newcommand{\Hm}{H_{-}}
\newcommand{\Hp}{H_{+}}
\newcommand{\QL}{Q_{\rm L}}
\newcommand{\QR}{Q_{\rm R}}
\newcommand{\vacL}{\langle 0\!\mid}
\newcommand{\vacR}{\mid \!\! 0 \rangle}

\newcommand{\Rpi}[1]{\mid \!\! \pi^{#1}\rangle}

\newcommand{\Mpic}{M_{\pi^{\scriptscriptstyle{+}}}}
\newcommand{\Mpin}{M_{\pi^{\scriptscriptstyle{0}}}}
\newcommand{\pin}{\pi^{\scriptscriptstyle{0}}}
\newcommand{\pic}{\pi^{\scriptscriptstyle{+}}}
\newcommand{\pim}{\pi^{\scriptscriptstyle{-}}}
\renewcommand{\vec}[1]{\textbf{#1}}
\newcommand{\tr}[1]{\langle #1 \rangle}
\newcommand{\Deltam}{m_d-m_u}
\newcommand{\Lagr}{\mathcal{L}}
\newcommand{\order}{\mathcal{O}}
\newcommand{\Mmatrix}{\mathcal{M}}
\newcommand{\nn}{\nonumber}
\newcommand{\lIR}{\lambda_{I\!R}}

\newcommand{\sss}{\scriptscriptstyle}
\begin{document}
\begin{titlepage}
\begin{flushright}BUTP-02/13\end{flushright}
\vspace*{1cm}
\begin{center}
  {\LARGE\bf Explicit generating functional for pions and virtual photons}

\vspace{1.8cm}
J. Schweizer\\Institute for Theoretical Physics,\\
University of Bern,\\ 
Sidlerstr. 5, CH--3012 Bern, Switzerland\\
E-mail: schweizer@itp.unibe.ch

\vspace{0.3cm}
March 31, 2003

\vspace{0.6cm}
\begin{abstract}
We construct the explicit one--loop functional of chiral perturbation theory
for two light flavours, including virtual photons. We stick to contributions
where 1 or 2 mesons and at most one photon are running in the loops. With
the explicit functional at hand,
the evaluation of the relevant Green functions boils down to performing
traces over the flavour matrices. For illustration, we work
out the $\pic\pim\rightarrow \pin\pin$ scattering amplitude
at threshold at order $p^4,e^2p^2$.
\end{abstract}
\vspace{2cm}
\footnotesize{\begin{tabular}{ll}
{{\bf{Pacs}} number(s):}$\!\!\!\!$&12.39.Fe, 13.40.Ks, 13.75.Lb\\
{\bf{Keywords:}}$\!\!\!\!$& Chiral Lagrangians, Pion, Isospin
symmetry breaking, \\& Electromagnetic corrections
\end{tabular}}
\end{center}
\end{titlepage}
\thispagestyle{empty}
\tableofcontents
\newpage

\setcounter{equation}{0}
\section{Introduction}
The low--energy structure of QCD may be investigated in the context of
chiral perturbation theory \cite{Weinberg, SU2, SU3}. In Refs. \cite{SU2, SU3}
Gasser and Leutwyler constructed the generating functional of ${\rm SU}(N_f)
\times {\rm SU}(N_f)$ for $N_f = 2, 3$ to order $p^4$. An explicit
representation of this functional, including graphs with up to two propagators
and at most four external pion fields\footnote{The formula presented also includes external vector, axial vector,
  scalar and pseudoscalar fields.},
was also given. For the first two flavours $u$ and $d$, this construction was
extended recently to include graphs with three propagators \cite{Unterdorfer}. 

In the case of QCD including electromagnetic interactions, the initial theory
depends on the strong coupling constant $g$, the fine structure constant $\alpha \simeq 1/137$ and the light quark masses. The corresponding effective theory was formulated in
Refs. \cite{resonance, Urech, Knecht&Urech, Meissner}. It is based on a systematic
expansion, which combines the chiral power counting scheme with the expansion
in powers of the electromagnetic coupling $e$. Within this framework, virtual
photon effects were calculated for a number of processes. In the two--flavour
case, electromagnetic corrections to the $\pi-\pi$ scattering amplitude
\cite{Knecht&Urech, Meissner, elpipiscattering} as well as to the vector and scalar form
factors \cite{piformfactor} have been evaluated at next-to-leading
order. Further, pionic beta decay and radiative $\tau$ decay have been analyzed
\cite{pibetadecay, taudecay} in a generalized framework including leptons and
virtual photons \cite{leptons}. Virtual photons have also been included in the
three--flavour case \cite{SU3em, K2pipi, weakem, KL3, piK}.

The purpose of the present article is to include virtual photons in the
explicit generating functional at one--loop \cite{SU2}. The advantage of
having the explicit functional at hand is evident: The calculation of S-matrix
elements boils down to performing traces over flavour matrices - the
combinatorics has already been carried out and all quantities are expressed in terms of
ultraviolet finite integrals. In the following, we stick to the two--flavour case. The extension to three flavours will be considered elsewhere \cite{SU3&photons}. 

The article is organized as follows: In the first part (Sects. \ref{slagr},
\ref{sZoneloop}), we construct the generating functional for ${\rm
  SU(2)_R} \times {\rm SU(2)_L} \times {\rm U(1)_V}$ to $\order{(p^4, e^2
  p^2)}$ in the low--energy expansion. This allows us to calculate Green functions to next-to-leading order
in a simultaneous expansion in powers of the external momenta, of the quark
masses and of the electromagnetic coupling. In order to extract form factors
or scattering amplitudes from our explicit representation of the generating
functional, one simply has to perform traces over flavour
matrices. The extraction procedure is demonstrated in Sects. \ref{sextract}
and \ref{spipi} by means of the $\pic\pim \rightarrow \pin\pin$ scattering
amplitude.

\setcounter{equation}{0}
\section{Lagrangian}
\label{slagr}
The variables of the effective theory are the pion field $U(x)\in {\rm SU}(2)$ and
the photon field $A_\mu(x)$. As shown in Ref. \cite{Urech}, the electric
charge must be - for consistency - treated as a quantity of order $p$ in the chiral expansion.
\subsection{Leading order}
At $\alpha\neq 0$, the leading order Lagrangian which is consistent with
chiral symmetry becomes \cite{resonance, Urech},
\begin{eqnarray}
  \Lagr^{(2)} &=& \frac{F^2}{4}\tr{d^\mu U^+ d_\mu U+\chi^+
 U+U^+\chi}-\frac{1}{4}F^{\mu\nu}F_{\mu\nu}\nn\\
&&- \frac{1}{2a}(\d^\mu A_\mu)^2+C\tr{\QR U \QL U^+},  
\label{Lagr2}              
\end{eqnarray}
with
\begin{equation}
  F_{\mu\nu} = \d_\mu A_\nu - \d_\nu A_\mu, \quad d_\mu U = \d_\mu U - i R_\mu U +i U L_\mu, \quad \chi = 2B(s+ip),
\end{equation}
and
\begin{eqnarray}
  R_\mu &=& v_\mu + A_\mu\QR + a_\mu, \nn\\ 
  L_\mu &=& v_\mu + A_\mu\QL - a_\mu.
\end{eqnarray}
The symbol $\tr{\dots}$ denotes the trace in flavour space. The external fields
$v_\mu$, $a_\mu$, $p$ and $s$ are given by
\begin{eqnarray}
  && v_\mu  = \tfrac{1}{2}v^0 {\bf 1}+ v^i\tfrac{\tau^i}{2}, \quad a_\mu  = a^i\tfrac{\tau^i}{2},\\\nn
  && s = s^0 {\bf 1}+ s^i\tau^i, \quad p = p^0 {\bf 1}+ p^i\tau^i,
\end{eqnarray}
where $\tau^i$ denote the Pauli matrices. We restrict ourselves to isovector axial fields, i.e. we take $\tr{a_\mu} = 0$.
For the transformation properties of the external fields, we refer to
\cite{SU2}. The mass matrix of the two light quarks is contained in $s$,
\begin{equation}
  s = \Mmatrix + \cdots, \quad \Mmatrix = {\rm diag}\left(m_u,m_d\right).
\end{equation}
To ensure the chiral symmetry of the effective Lagrangian (\ref{Lagr2}), the 
local right-- and left--handed spurions $\QR$ and $\QL$ transform under
${\rm SU}(2)_{\rm R} \times {\rm SU}(2)_{\rm L}$, according to
\begin{equation}
  Q_{\rm I} = V_{\rm I} Q_{\rm I} V_{\rm I}^+, \quad {\rm I} = {\rm R, L}.
\end{equation}
In the following, we work with a constant charge matrix
\begin{equation}
  \QR = \QL = Q = \frac{e}{3}{\rm diag}\left(2,-1\right).
\end{equation} 
The quantity $a$ denotes the gauge fixing parameter and the
parameters $F$, $B$ and $C$ are the three low--energy coupling constants occurring at
leading order. In the chiral limit and in absence of electromagnetic
interactions, $F$ coincides with the pion decay constant, normalized such that
$F_\pi = 92.4 {\rm MeV}$ \cite{pdg}, while $B$ is related
to the quark condensate. The coupling constant $C$ occurs in the low--energy expansion of the charged physical pion mass only. 

Finally, the matrix $\bar{U}$ is determined by the equation of motion
 \begin{eqnarray}
   d_\mu d^\mu \bar{U}\bar{U}^+-\bar{U}d_\mu
   d^\mu\bar{U}^++\bar{U}\chi^+-\chi\bar{U}^+-\frac{1}{2}\langle \bar{U}\chi^+-\chi\bar{U}^+\rangle &&\nn\\
+\frac{4C}{F^2}\left(\bar{U}Q\bar{U}^+ Q-Q\bar{U}Q\bar{U}^+\right) &=& 0,
\label{cleq}
 \end{eqnarray}
while the field equations for the photon field $\bar{A}_\mu$ read
\begin{equation}
  \left[ g_{\mu\nu} \Box-\left(1-\tfrac{1}{a}\right)\d_\mu
  \d_\nu\right]\bar{A}^\nu +\frac{i F^2}{2}\tr{d_\mu
  \bar{U}[\bar{U}^+ ,Q]} = 0.
\label{cleqA}
\end{equation}
\subsection{Next-to-leading order}
The next-to-leading order Lagrangian reads
\begin{equation}
 \Lagr^{(4)} =  \Lagr_{p^4}+\Lagr_{p^2e^2}+\Lagr_{e^4}.
\label{Lagr4}
\end{equation}
The Lagrangian at
order $p^4$ was constructed in Refs. \cite{SU2, Knecht&Urech, Kaiser}\footnote{Ref. \cite{Knecht&Urech} uses a different convention for the
coupling constant $h_2$.},
\begin{eqnarray}
\Lagr_{p^4} &=& \frac{l_1}{4}\langle d^{\mu}U^+ d_{\mu}U\rangle^2 
+ \frac{l_2}{4}\langle d^{\mu}U^+ d^{\nu}U\rangle\langle
d_{\mu}U^+ d_{\nu}U\rangle + \frac{l_3}{16}\langle\chi^+ U+U^+ \chi\rangle^2\nn\\
&& +\frac{l_4}{4}\langle d^{\mu}U^+ d_{\mu}\chi + d^{\mu}\chi^+
d_{\mu}U\rangle + l_5\langle \hat{R}_{\mu\nu}U \hat{L}^{\mu\nu} U^+\rangle\nn\\
&& +\frac{i l_6}{2}\langle \hat{R}_{\mu\nu}d^{\mu}U d^{\nu}U^+ 
+ \hat{L}_{\mu\nu}d^{\mu}U^+ d^{\nu}U\rangle - \frac{l_7}{16}\langle\chi^+ U-U^+\chi\rangle^2\nn\\
&& + \frac{1}{4}(h_1+h_3)\langle \chi^+\chi\rangle +
\frac{1}{2}(h_1-h_3) {\rm Re} ({\rm det}\chi)\nn\\
&& -\frac{1}{2}(l_5+4h_2)\langle
\hat{R}_{\mu\nu}\hat{R}^{\mu\nu}+\hat{L}_{\mu\nu}\hat{L}^{\mu\nu}\rangle\nn\\
&& +\frac{h_4}{4}\tr{R_{\mu\nu}+L_{\mu\nu}}\tr{R^{\mu\nu}+L^{\mu\nu}},
\end{eqnarray}
with right-- and left--handed fields strengths defined as
\begin{eqnarray}
 && R_{\mu\nu} = \d_\mu
  R_\nu-\d_\nu R_\mu-i\left[R_\mu, R_\nu\right], \quad \hat{R}_{\mu\nu} =
  R_{\mu\nu}-\frac{1}{2}\tr{R_{\mu\nu}},\nn\\
&& L_{\mu\nu} = \d_\mu
  L_\nu-\d_\nu L_\mu-i\left[L_\mu, L_\nu\right], \quad \hat{L}_{\mu\nu} =
  L_{\mu\nu}-\frac{1}{2}\tr{L_{\mu\nu}}.
\end{eqnarray}
The most general list of counterterms occurring at order $p^2 e^2$ was given
in Refs. \cite{Knecht&Urech, Meissner},
\begin{eqnarray}
\Lagr_{p^2e^2} &=& F^2 \big\{ k_1\langle d^{\mu}U^+ d_{\mu} U\rangle
\langle Q^2 \rangle +k_2 \langle d^{\mu}U^+ d_{\mu}U\rangle\langle 
QUQU^+ \rangle 
\nn\\
&&+ k_3\big(\langle d^{\mu}U^+ Q U \rangle\langle d_{\mu}U^+ Q U 
\rangle + \langle d^{\mu}U Q U^+ \rangle\langle
d_{\mu}U Q U^+\rangle\big)
\nn\\
&&+ k_4\langle d^{\mu}U^+ Q U\rangle\langle d_{\mu}U Q U^+ \rangle
+k_5\langle\chi^+ U+U^+ \chi\rangle\langle Q^2\rangle\nn\\
&&+ k_6\langle\chi^+ U+U^+\chi\rangle\langle Q U Q U^+ \rangle
\nn\\
&&+ k_7\langle (\chi U^+ + U\chi^+ )Q+
(\chi^+ U+U^+\chi )Q\rangle\langle Q\rangle
\nn\\
&&+ k_8\langle(\chi U^+ - U\chi^+ )QUQU^+
+(\chi^+ U-U^+\chi )Q U^+ Q U\rangle
\nn\\
&&+ k_9 \langle d_{\mu}U^+ [c_{\rm R}^{\mu}Q,Q]U
+d_{\mu}U[c_{\rm L}^{\mu}Q,Q]U^+\rangle
\nn\\
&&+ k_{10}\langle c_{\rm R}^{\mu}Q U c_{\rm L \mu}Q U^+\rangle +
k_{11}\langle c_{\rm R}^\mu Q c_{\rm R \mu} Q + c_{\rm L}^\mu Q  
c_{\rm L\mu}Q \rangle\big\},
\end{eqnarray}
where
\begin{equation}
 c^\mu_{\rm I} Q = -i[I^\mu,Q], \quad I = R,L.
\end{equation}
In the following we consider the next-to-leading order contributions
where at most one virtual photon is running in a loop. Therefore, we
drop the term $\Lagr_{e^4}$ in Eq.~(\ref{Lagr4}), (terms with $k_{12},
k_{13}$ and $k_{14}$ in Ref.~\cite{Knecht&Urech}).

The renormalized couplings are defined by 
\begin{eqnarray}
  l_i &=& l^r_i +\gamma_i \lambda, \nn\\
  h_i &=& h^r_i +\delta_i \lambda, \nn\\
  k_i &=& k^r_i +\sigma_i \lambda,
\end{eqnarray}
with
\begin{equation}
  \lambda = \frac{\mu^{d-4}}{16\pi^2}\left[\frac{1}{d-4}-\frac{1}{2}\left({\rm
  ln}4\pi +\Gamma'(1)+1\right)\right].
\end{equation} 
The coupling constant $h_4$ is finite at $d = 4$ \cite{Kaiser}, because the
singlet fields $\tr{v_\mu}$ and $A_\mu \tr{Q}$ do not occur in the leading
order Lagrangian (\ref{Lagr2}). The coefficients $\gamma_i, \delta_i$ and
$\sigma_i$ are specified in Refs. \cite{SU2} and \cite{Knecht&Urech}.

\setcounter{equation}{0}
\section{Generating functional at one--loop}
\label{sZoneloop}
To evaluate the one--loop contribution to the generating functional, we expand
$U$ and $A_\mu$ around their solutions
$\bar{U}=u^2$ and $\bar{A}_\mu$ to the classical equations of motion,
\begin{eqnarray}
  U &=& u(1+i\xi-\tfrac{1}{2}\xi^2+\cdots)u, \quad \xi = \frac{1}{F}\xi^i
 \tau^i, \nn\\
 A_\mu &=& \bar{A}_\mu+\epsilon_\mu.
\end{eqnarray}
Collecting the fluctuations in $\eta=(\xi^1, \dots, \xi^3, \epsilon^1, \dots,
\epsilon^4)$, the Euclidean action can be written as a quadratic form \cite{Urech, Knecht&Urech},
\begin{equation}
  \int dx_E \Lagr^{(2)}_E = \int dx_E \bar{\Lagr}^{(2)}_E + \tfrac{1}{2}\int dx_E
  \eta^A D^{AB}_E \eta^B.
\end{equation}
Here, $\bar{\Lagr}^{(2)}_E$ is evaluated at the solution to the classical equations of
motion. The one--loop functional $ Z_{E{\rm one loop}}$ then takes the form:
\begin{equation}
  Z_{E{\rm one loop}} = \frac{1}{2}{\rm ln \,det}D_E.
\end{equation}
\subsection{Differential operator}
In the Feynman gauge $a=1$, the Euclidean differential operator is given by
\begin{equation}
  D^{AB}_E = -(\Sigma_\mu \Sigma_\mu)^{AB}+\Lambda^{AB} \quad A,B = 1,
  \dots, 7,
\end{equation}
with 
\begin{equation}
   \Sigma_\mu = \d_\mu \op{1}+ Y_\mu, \quad Y_\mu= \left( \begin{array}{cc}
  \Gamma_\mu^{ik} & X_\mu^{i\rho} \\ X_\mu^{\sigma k} & 0
  \end{array}\right),\quad \Lambda = \left( \begin{array}{cc} \sigma^{ik} &
  \tfrac{1}{2}\gamma^{i\rho}\\ \tfrac{1}{2}\gamma^{\sigma k} & \rho
  \delta^{\sigma\rho} \end{array}\right).
\end{equation}
The pion field indices $i, k$ run from $1$ to $3$, while the photon field
components are labeled by Greek letters $\sigma,\rho = 1, \dots, 4$. 

To renormalize the determinant we work in $d\neq4$ dimensions,
\begin{eqnarray}
  \sigma^{ik} &=&
  -\frac{1}{2}\tr{[\Delta_\mu,\tau^i][\Delta_\mu,\tau^k]}+\frac{1}{2}\delta^{ik}\tr{\sigma}-\frac{d F^2}{16}\tr{\Hm\tau^i}\tr{\Hm\tau^k}\nn\\
&&
  -\frac{C}{8F^2}\tr{[\Hp\!+\!\Hm,\tau^i][\Hp\!-\!\Hm,\tau^k]+i\leftrightarrow k},\nn\\
\Gamma^{ik}_\mu &=& -\frac{1}{2}\tr{[\tau^i,\tau^k]\Gamma_\mu},\nn\\
\gamma^i_\mu &=&
  F\tr{([\Hp,\Delta_\mu]+\frac{1}{2}D_\mu\Hm)\tau^i},\nn\\
 X^{i\rho}_\mu &=& -X^{\rho i}_\mu =  -\frac{F}{4}\tr{\Hm\tau^i}\delta^\rho_\mu,\nn\\
\rho &=& \frac{3F^2}{8}\tr{\Hm^2},
\end{eqnarray}
where 
\begin{eqnarray}
  D_\mu \Hm &=& \d_\mu \Hm +\left[\Gamma_\mu,\Hm\right],\nn \\
 \Gamma_\mu &=& \frac{1}{2}\left[u^+,\d_\mu u
 \right]-\frac{i}{2}u^+(v_\mu+\bar{A}_\mu \QR+a_\mu)
  u\nn\\
&&-\frac{i}{2}u(v_\mu+\bar{A}_\mu \QL-a_\mu)u^+,
\end{eqnarray}
and
\begin{eqnarray}
  \Delta_\mu &=& \frac{1}{2}u^+ d_\mu \bar{U} u^+ =
  -\frac{1}{2}ud_\mu \bar{U}^+ u,\nn\\
\sigma &=& \frac{1}{2}\left(u^+\chi u^++u\chi^+ u\right),\nn\\
\Hp &=& u^+ \QR u + u\QL u^+,\nn\\
\Hm &=& u^+ \QR u - u\QL u^+.
\end{eqnarray}
In the absence of external fields $D_{E}$ reduces to
\begin{equation}
  D_{0 E} = \left(\begin{array}{cc}(-\d_\mu\d_\mu+M_i^2)\delta^{ik} & \\ 
 & -\d_\mu\d_\mu\delta^{\sigma\rho}\end{array}\right).
\end{equation}
The parameters $M_i$ denote the pion masses at leading order  
\begin{equation}
  M_1^2 = M_2^2 = M_+^2 = 2B \hat{m}+\frac{2Ce^2}{F^2}, \quad M_3^2 = M_0^2 = 2B \hat{m},
\end{equation}
with $\hat{m} = (m_u+m_d)/2$. The full operator is given by
\begin{eqnarray}
  D_E &=& D_{0 E}+\delta_E, \nn\\
  \delta_E &=& -\{ \d_\mu,Y_\mu \}-Y_\mu Y_\mu+\bar{\Lambda},
\end{eqnarray}
where 
\begin{equation}
  \bar{\Lambda} = \left( \begin{array}{cc} \bar{\sigma}^{ik} &
  \tfrac{1}{2}\gamma^{i\rho}\\ \tfrac{1}{2}\gamma^{\sigma k} & \rho
  \delta^{\sigma\rho}\end{array}\right), \quad \bar{\sigma}^{ik} = \sigma^{ik}-M_i^2\delta^{ik}.
\end{equation}
\subsection[Expansion in $\delta_E$]{Expansion in \boldmath{$\delta_E$}}
To work out explicitly the one--loop contributions to a given Green function, we
have to expand the determinant of the differential operator in powers of the
external fields. Since $\delta_E$ vanishes if the external fields are switched
off, we may expand ${\rm ln\,det}D_E$ in powers of $\delta_E$,
\begin{eqnarray}
  Z_{E\,{\rm one loop}}&=&\tfrac{1}{2}\,{\rm ln\,det}D_{0 E} +\tfrac{1}{2}\langle
  D_{0 E}^{-1}\delta_E\rangle-\tfrac{1}{4}\langle D_{0 E}^{-1}\delta_E
  D_{0 E}^{-1}\delta_E\rangle\nn\\
&& +\tfrac{1}{6}\langle D_{0 E}^{-1}\delta_E D_{0 E}^{-1}\delta_E
  D_{0 E}^{-1}\delta_E \rangle+\cdots.
\label{expdet}
\end{eqnarray}
Following the counting scheme used in Refs. \cite{SU2,SU3}, the external fields
$a_\mu$ and $p$ count as $\order{(\Phi)}$, whereas $v$ and $s\!-\!\Mmatrix$ are of $\order(\Phi^2)$. 

In the presence of electromagnetic interactions $\delta_E$ is of $\order(\Phi)$
rather than $\order(\Phi^2)$, since the quantities $\gamma$ and $X$ count as
$\order(\Phi)$. To achieve an accuracy of $\order{(\Phi^4)}$, it is sufficient
to stop the expansion in (\ref{expdet}) at $\order{(\delta_E^3)}$. This is due to
the fact that both, $\gamma$ and $X$, are of order $e$, and their
contributions at $\order{(\Phi^4)}$ are beyond the precision of our
calculation.
\subsection[Tadpole and unitarity contributions to order $\Phi^4$]{Tadpole and unitarity contributions to order \boldmath{$\Phi^4$}}
\label{sexplZ}
\begin{figure}[htbp]
\begin{center}
    \makebox{\includegraphics[height=2.0cm]{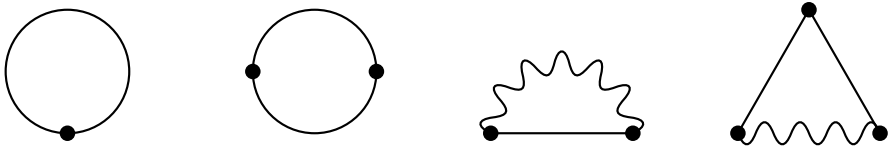}}
\caption{One--loop diagrams: The dots denote polynomials in the classical
    fields $\bar{U}$, $\bar{A}_\mu$, as well as in the external fields $v_\mu$,
    $a_\mu$, $p$ and $s$. The plain line stands for the pion propagator and 
    the wiggled line describes the photon.}
    \label{fig: diagrams}
  \end{center}
\end{figure}

In the following, we work in Minkowski spacetime.
If we add $\int dx (\bar{\Lagr}^{(2)}+\bar{\Lagr}^{(4)})$ to $Z_{\rm one loop}$ and
renormalize the low--energy couplings according to \cite{SU2, Knecht&Urech}, the
result for the generating functional to $\order{(p^4, e^2p^2)}$ is ultraviolet
finite as $d\rightarrow4$. At order $\Phi^4$, it takes the form, (see figure
\ref{fig: diagrams})
\begin{equation}
  Z = Z_t + Z_u.
\end{equation}
Explicitly, $Z_t$ consists of
\begin{eqnarray}
Z_t &=& -\frac{1}{32\pi^2} M_i^2 {\rm ln}\frac{M_i^2}{\mu^2}\int dx\,
\bar{\sigma}_{ii}(x)+\frac{1}{16\pi^2}M_i^2 \int dx Y_i(x)Y_i(x)\nn\\
&& +\frac{1}{16\pi^2}\int dx\,\bar{\sigma}_{ik}(x)Y_k(x)Y_i(x)-
\frac{1}{48\pi^2}\int dx D_\mu Y_i(x) D^\mu Y_i(x) \nn\\
&& +\int dx\,\bar{\Lagr}_2(x) +\int dx\,\bar{\Lagr}_4^r(x),
\label{Zt}
\end{eqnarray}
where the first term contains tadpole contributions, the next three terms
are generated by expanding $Z_u$ around $d=4$, while the last two terms are tree
graphs. Repeated isospin indices are summed over. The quantity $\bar{\sigma}_{ik}$ now corresponds to\footnote{For further convenience, we specify $\bar{\sigma}_{ik}$ in $d \neq
  4$ dimensions. However, Eq. (\ref{Zt}) is both, ultraviolet and infrared
  finite as $d\rightarrow 4$.}
\begin{eqnarray}
  \bar{\sigma}_{ik} &=&
  \frac{1}{2}\tr{\left[\Delta_\mu,\tau_i\right]\left[\Delta^\mu,\tau_k\right]}+\frac{1}{2}\delta_{ik}\tr{\sigma}-\frac{d F^2}{16}\tr{\Hm\tau_i}\tr{\Hm\tau_k}\nn\\
&&
  -\frac{C}{8F^2}\tr{\left[\Hp\!+\!\Hm,\tau_i\right]\left[\Hp\!-\!\Hm,\tau_k\right]+i\leftrightarrow k}- M_i^2\delta_{ik},
\end{eqnarray}
and 
\begin{equation}
  Y_i = \frac{F}{4}\tr{\Hm\tau_i}, \quad D^\mu Y_i = \d^\mu Y_i +\Gamma_{ik}^\mu Y_k.
\end{equation}
The unitarity correction $Z_u = Z_{u_2} + Z_{u_3}$ contains one--loop graphs with two vertices,
\begin{eqnarray}
Z_{u_2} &=& \int dx
dy\,\big\{M_{\mu\nu}^r(x\!-\!y,M_i^2,M_k^2)\Gamma^\mu_{ik}(x)\Gamma^\nu_{ki}(y)\nn\\
&& +2 M_{\mu}^{\mu r}(x\!-\!y,M_i^2,0) Y_i(x) Y_i(y)\nn\\
&& +K_\mu(x\!-\!y,M_i^2,M_k^2)\left[\Gamma^\mu_{ik}(x)\bar{\sigma}_{ki}(y)+d \Gamma^\mu_{ik}(x)Y_k(y)Y_i(y)\right]
\nn\\
&& -K_\mu(x\!-\!y,M_i^2,0)\left[Y_i(x)\gamma_i^\mu(y)+2Y_i(x)\Gamma^{\mu}_{ik}(y)Y_k(y)\right]
\nn\\
&&
+\tfrac{1}{4}J^r(x\!-\!y,M_i^2,M_k^2)\left[\bar{\sigma}_{ik}(x)\bar{\sigma}_{ki}(y)+2d
  \bar{\sigma}_{ik}(x)Y_k(y)Y_i(y)\right]
\nn\\
&&
-\tfrac{1}{8}J^r(x\!-\!y,M_i^2,0)\left[\gamma^\mu_i(x)\gamma_{i\mu}(y)+4\gamma_{i\mu}(x)\Gamma^\mu_{ik}(y)Y_k(y)\right]\big\},
\label{Zu2}
\end{eqnarray}
as well as the triangle vertex diagram with two pion and one photon
propagator,
\begin{eqnarray}
Z_{u_3} &=& \int dx dy dz \, \big\{\tfrac{1}{8}
T(x\!-\!y,y\!-\!z, M_i^2, M_k^2)\gamma^\mu_i(x)\bar{\sigma}_{ik}(y)\gamma_{k\mu}(z)
\nn\\
&&
+\tfrac{1}{4}T_{1\mu}(x\!-\!y,y\!-\!z, M_i^2, M_k^2)Y_i(x)\bar{\sigma}_{ik}(y)\gamma_k^\mu(z)
\nn\\
&& -\tfrac{1}{4}T_{2\mu}(x\!-\!y,y\!-\!z, M_i^2, M_k^2)\gamma^\mu_i(x)\bar{\sigma}_{ik}(y)Y_k(z)
\nn\\
&&
-\tfrac{1}{8}T_{3\mu}(x\!-\!y,y\!-\!z, M_i^2, M_k^2)\gamma^\nu_i(x)\Gamma^\mu_{ik}(y)\gamma_{k\nu}(z)
\nn\\
&&+\tfrac{1}{4}T_{1\mu\nu}^r(x\!-\!y,y\!-\!z, M_i^2, M_k^2)Y_i(x)\Gamma^\mu_{ik}(y)\gamma_k^\nu(z)
\nn\\
&&-\tfrac{1}{4}T_{2\mu\nu}^r(x\!-\!y,y\!-\!z, M_i^2, M_k^2)\gamma^\mu_i(x)\Gamma^\nu_{ik}(y)Y_k(z)
\nn\\
&&+\tfrac{1}{2}T_{3\mu}^{\mu r}(x\!-\!y,y\!-\!z, M_i^2, M_k^2)Y_i(x)\bar{\sigma}_{ik}(y)Y_k(z)
\nn\\
&&-\tfrac{1}{2}T_{\mu\nu}^{\mu r}(x\!-\!y,y\!-\!z, M_i^2, M_k^2)Y_i(x)\Gamma^\nu_{ik}(y)Y_k(z)\big\}.
\label{Zu3}
\end{eqnarray} 
For the definition of the various kernels, we refer the reader to appendix
\ref{Kernels}. For non-exceptional momenta, the generating
functional (\ref{Zt}), (\ref{Zu2}) and (\ref{Zu3}) is finite as
$d\rightarrow 4$. However, we want to extract on--shell matrix elements and
this leads in general to infrared singularities. We treat these infrared pole
terms using dimensional regularization. Both, the kernels as well as the
flavour matrices are thus written in $d \neq 4$ dimensions. 

In the isospin symmetry limit $m_u = m_d$ and $\alpha=0$, our expression
agrees with the explicit one--loop functional of ${\rm SU}(2)\times {\rm
  SU}(2)$ in Ref. \cite{SU2}.
  
The representation for the generating functional, specified in (\ref{Zt}),
(\ref{Zu2}) and (\ref{Zu3}), contains the one--loop contributions to Green
functions in the even intrinsic parity sector, formed with the quark currents,
\begin{equation}
  S^i =\bar{q}\tau^i q, \quad S^0 = \bar{q}q, \quad P^i
  =\bar{q}i\gamma_5\tau^i q, \quad P^0 = \bar{q}i\gamma_5q,
\end{equation}
and 
\begin{equation}
V_\mu^i =
  \frac{1}{2}\bar{q}\gamma_\mu \tau^i q, \quad V_\mu^0 =
  \frac{1}{2}\bar{q}\gamma_\mu q, \quad
  A_\mu^i = \frac{1}{2}\bar{q}\gamma_\mu \gamma_5 \tau^i q.
\end{equation}
The accuracy $\order{(\Phi^4)}$ suffices to calculate all two--point functions,
the three--point functions containing one vector or scalar current, as
well as the pseudoscalar and axial four--point functions. 

The calculation of form factors or scattering amplitudes from the explicit
representation of the generating functional (\ref{Zt}),
(\ref{Zu2}) and (\ref{Zu3}) amounts to perform the traces over the flavour matrices. In Sect. \ref{sextract}, we demonstrate the extraction
procedure by means of the $\pi-\pi$ scattering amplitude.

\setcounter{equation}{0}
\section{Masses and decay constants}
In the following, it is convenient to work in $\sigma$-parameterization,
i.e. to represent the matrix $U(x)$ in the form
\begin{equation}
  U(x) = \sqrt{1-\frac{\tr{\phi^2}}{2}}+i\phi,\quad \phi = \frac{1}{F}\phi^i
  \tau^i,
\end{equation}
however, all the steps listed are parameterization independent. 
\subsection[Generating functional at order $\Phi^2$]{Generating functional at
  order \boldmath{$\Phi^2$}}
\label{Normalization}
We expand the generating functional in powers of $\Phi$,
\begin{equation}
  Z = {\rm const}+ Z^{(2)}+Z^{(4)}+\cdots,
\end{equation}
where the pion field $\bar{\phi}$ counts as $\order{(\Phi)}$ and the photon field $\bar{A}_\mu$ is of order $\Phi^2$. This can be seen by
solving the equations of motion (\ref{cleq}) and (\ref{cleqA}) to first and second order in $\Phi$, respectively. The field equations for the pion field are discussed in subsection \ref{eom}.

In the presence of electromagnetic interactions, the quadratic term $Z^{(2)}$
contains non-local contributions coming from $Z_{u_2}$. If we expand their
Fourier transform around the physical pion masses $M_{\pi_i}$, we arrive at 
\begin{eqnarray}
  Z^{(2)}\left(p, a_\mu, \bar{\phi}\right) &=& \int dx-\frac{1}{2}Z_i^{-1}\bar{\phi}^i\left(\Box+M_{\pi_i}^2\right)\bar{\phi}^i
  +Z_i^{-\frac{1}{2}}F_{\pi_i} \d^\mu a^i_\mu\bar{\phi}^{i}\nn\\
&& +Z_i^{-\frac{1}{2}}G_{\pi_i} p^i\bar{\phi}^{i}+Z_3^{-\frac{1}{2}}\tilde{G}_{\pi} p^0\bar{\phi}^{3}(x)+\cdots.
\label{Zphi2}
\end{eqnarray}
The ellipsis stands for terms which do not contribute to the residue of the pole in the axial or pseudoscalar two--point functions.

The scaling factors $Z_0 = Z_3$ and $Z_+ = Z_1 = Z_2$ are given by
\begin{eqnarray}
  Z_0 &=& 
  1-\frac{20e^2}{9}\left[k^r_1+k^r_2+\frac{9}{10}\left(k^r_4-2k^r_3\right)\right]+\frac{M_+^2}{8\pi^2F^2}{\rm ln} \frac{M_+^2}{\mu^2},\nn\\
Z_+ &=&  1+4e^2\lIR-\frac{20e^2}{9}\left(k^r_1+k^r_2\right)\nn\\
&&+\frac{1}{16\pi^2 F^2}\left[M_+^2{\rm ln}\frac{M_+^2}{\mu^2}+M_0^2{\rm
  ln}\frac{M_0^2}{\mu^2}\right].
\label{Zfactor}
\end{eqnarray}
We used dimensional regularization to treat the infrared singularity,
\begin{equation}
  \lIR = \frac{\mu^{d-4}}{16\pi^2}\left[\frac{1}{d-4}-\frac{1}{2}\left({\rm
  ln}
  4\pi +\Gamma'(1)+1\right)\right].
\end{equation}
Note that the scaling factors $Z_0$ and $Z_+$ are parameterization
independent, since unitarity determines the expansion of the field $U$ to $\order{(\Phi^2)}$. 
\subsection{Physical masses}
We choose the charged pion mass as normalization point \cite{pipiatom}. In the
isospin-symmetry limit $m_u = m_d$ and $\alpha = 0$, the position of the pole
in the correlator of two axial currents is identified with the charged pion
mass. The physical masses then read at next-to-leading order\footnote{The
  coefficient of $(\Deltam) \, k_7$ differs by a factor $\frac{1}{2}$ from
  the corresponding term obtained in Ref. \cite{Knecht&Urech}.},
\begin{eqnarray}
\Mpic^2 &=& M_+^2+\Mpic^2\bigg\{\frac{e^2}{4\pi^2}-\frac{\Mpic^2-4e^2F^2Z}{32\pi^2F^2}\bar{l}_3+\frac{e^2}{32\pi^2}\left[\left(3+\tfrac{4Z}{9}\right)\bar{k}_1-\tfrac{40Z}{9}\bar{k}_2\right.\nn\\
&& \left.-\tfrac{1}{9}\left(5+4Z\right)\bar{k}_5+\tfrac{23}{9}\left(1+8Z\right)\bar{k}_6+\left(1-8Z\right)\bar{k}_8\right]+\frac{4e^2}{9}k_7\nn\\
&&-\frac{\Mpic^2-4e^2F^2Z}{32\pi^2F^2}{\rm
  ln }\frac{\Mpic^2}{\Mpin^2}\bigg\}-\frac{4e^2B(\Deltam)}{3}k_7+\order{(e^4,
  p^6)},
\label{Mcphys}
\end{eqnarray}
and
\begin{eqnarray}
  \lefteqn{\Delta_\pi = \Mpic^2-\Mpin^2=}\nn\\
&& 2e^2F^2Z + \frac{\Mpic^2}{32\pi^2}\bigg\{ e^2\left[ 8+ 3\bar{k}_3+4Z
  \bar{k}_4+2\left(1+8Z\right)\bar{k}_6+\left(1-8Z\right)\bar{k}_8\right]\nn\\
&& -\frac{2}{F^2}(\Mpic^2-4e^2F^2Z){\rm
  ln }\frac{\Mpic^2}{\Mpin^2}\bigg\}+\frac{2B^2(\Deltam)^2}{F^2}l_7+\order{(e^4,
  p^6)},
\label{DeltaPi}
\end{eqnarray}
with $Z \doteq C/F^4$. The scale independent low--energy constants $\bar{l}_i$ and $\bar{k}_i$ are defined by 
\begin{equation}
 l_i^r = \frac{\gamma_i}{32\pi^2}\big(\bar{l_i}+{\rm
 ln}\frac{\Mpic^2}{\mu^2}\big),\quad k_i^r = \frac{\sigma_i}{32\pi^2}\big(\bar{k_i}+{\rm
 ln}\frac{\Mpic^2}{\mu^2}\big).
\end{equation}
\subsection[Coupling constants $F_{\pi}$ and $G_{\pi}$
 at $\alpha \neq 0$]{Coupling constants \boldmath{$F_{\pi}$} and \boldmath{$G_{\pi}$}
 at \boldmath{$\alpha \neq 0$}}
The quantities $F_{\pi_i}$ and $G_{\pi_i}$ are the coupling constants of the
isovector axial and pseudoscalar currents to the pion,
\begin{equation}
  \vacL A^i_\mu \Rpi{k} =
  i p_\mu\delta^{ik}F_{\pi_i}, \quad \vacL P^i \Rpi{k} = \delta^{ik} G_{\pi_i},
\end{equation}
while $\tilde{G}_\pi$ stands for the one-particle matrix element of the
isoscalar density,
\begin{equation}
  \vacL P^0 \Rpi{3} = \tilde{G}_{\pi}.
\end{equation}
The coupling constants $F_{\pi_i}$, $G_{\pi_i}$ and $\tilde{G}_\pi$ are given
explicitly in appendix \ref{decayconstants}. 

For a constant charge matrix $Q$ the generating functional is invariant
under the subgroup of ${\rm SU(2)_{R}}\times{\rm SU(2)_{L}}$ transformations
\begin{equation}
  V_{\rm R} = {\rm exp}(i a \tau^3), \quad V_{\rm L} = {\rm exp}(i b \tau^3).
\end{equation}
This implies in particular that the relation,
\begin{equation}
  F_{\pin} \Mpin^2 = \hat{m} G_{\pin} + \frac{m_u-m_d}{2}\tilde{G}_\pi,
\end{equation}
holds at $\alpha \neq 0$. We checked that the
explicit expressions for $\Mpin^2, F_{\pin}$, $G_{\pin}$ and $\tilde{G}_\pi$
satisfy this relation to order $p^4, p^2e^2$.

\setcounter{equation}{0}
\section{Extracting the \boldmath{$\pic\pim \rightarrow \pin\pin$} amplitude
  from the generating functional}
\label{sextract}
To extract the scattering amplitude from the generating functional, we may
concentrate on the pseudoscalar four--point function. The external fields
$a_\mu$, $v_\mu$ and $p^0$ are switched off and the scalar field $s$ is set to
$\Mmatrix$. We may even switch $\bar{A}_\mu$ off, because there are no
one-photon exchange contributions to $\pic\pim \rightarrow \pin\pin$ at tree level. Now $Z = Z(p^i, \bar{\phi})$
is a functional of $p^i$ only, depending on $p^i$ both explicitly and implicitly through the equation of motion (\ref{cleq}). 

The pseudoscalar four--point function is given by the coefficient of $p^{i_1},
\dots, p^{i_4}$ in the expansion of $Z(p^i, \bar{\phi})$ in
powers of $\Phi$. 
\subsection{Equation of motion}
\label{eom}
The solution of the equation of motion (\ref{cleq}) represents a power series
in $\Phi$,
\begin{equation}
  \bar{\phi}= \phi^{(1)}+\phi^{(3)}+\cdots.
\end{equation}
To order $\Phi^4$, only the leading order term gives a contribution to the
action. Explicitly, $\phi^{(1)}$ reads
\begin{equation}
  \phi^{(1)i}(x)=2B F\int dy \Delta_i(x-y) p^i(y),
\end{equation}
where $\Delta_{i}$ stands for the Feynman propagator of a pion with mass $M_i$. Since $\bar{\phi}$ represents an extremum of the action $\int dx
\Lagr^{(2)}$, the classical solution may be replaced with the extremum of (\ref{Zphi2}),
\begin{eqnarray}
 \hat{\phi}^i(x)&=& Z_i^{\frac{1}{2}}G_{\pi_i}\int dy
 \Delta_{\pi_i}(x-y)p^i(y)+\order{(\Phi^3)},\nn\\
 \hat{\phi}(x)&=& \bar{\phi}(x) + \order{(p^2)},
\end{eqnarray}
where $\Delta_{\pi_i}$ denotes the Feynman propagator with physical mass $M_{\pi_i}$. The shift $\bar{\phi} \rightarrow \hat{\phi}$ affects the classical
action at order $p^6$ only,  
\begin{equation}
  \int dx \Lagr^{(2)}\{ \hat{\phi}\} = \int dx \Lagr^{(2)}\{ \bar{\phi}\}+\order{(p^6)},
\end{equation}
which is beyond one--loop accuracy.
\subsection{Recipe}
We may determine the scattering matrix element from the on--shell residue of
the fourfold pole in the Fourier transform of the pseudoscalar four--point function,
\begin{eqnarray}
  &&\!\!i^4\!\int dx_1 \dots dx_4 e^{i(p_3 x_3+p_4x_4-p_1x_1-p_2x_2)}\vacL T
  P^i(x_1)P^k(x_2)P^l(x_3)P^m(x_4)\vacR \nn\\
&&\!\! = \frac{G_{\pi_i}G_{\pi_k}G_{\pi_l}G_{\pi_m}}{(M_{\pi_i}^2-p_1^2)\dots(M_{\pi_m}^2-p_4^2)}\langle\pi^l(p_3)\pi^m(p_4)
  {\rm \sss out}\!\mid \!\pi^i(p_1)\pi^k(p_2){\rm \sss in}\rangle_c+\cdots.
\label{fourpoint}
\end{eqnarray}
The amplitude is related to the connected part of the matrix element through
\begin{equation}
  \langle\pi^l(p_3)\pi^m(p_4)
  {\rm \sss out}\!\mid \!\pi^i(p_1)\pi^k(p_2){\rm \sss in}\rangle_c = i(2\pi)^4
  \delta^4(p_1\!+\!p_2\!-\!p_3\!-\!p_4) A^{ik;lm}(s,t,u).
\end{equation}
Subsequently, we present a simple recipe to extract the $\pi^1\pi^1\rightarrow
\pi^3\pi^3$ amplitude from the
explicit representation of the generating functional (\ref{Zt}), (\ref{Zu2}) and (\ref{Zu3}):
\begin{itemize}
\item[-]
Switch all explicit external fields off, including the field $p^i$. Now
the generating functional $Z=Z(0, \hat{\phi})$ depends on $\hat{\phi}$ only. 
\item[-]
Perform the traces over the flavour matrices $\sigma$, $\Gamma_\mu\Gamma_\nu$, $\gamma_\mu \gamma^\mu, \dots$
\item[-]
The pion fields are rescaled, according to $\phi^{r i} =
Z_i^{-\frac{1}{2}}\hat{\phi}^i$. This step brings the quadratic terms
(\ref{Zphi2}) to normal form.
\item[-]
The scattering matrix element may be read off from the terms of order
$(\phi^{r 1})^2(\phi^{r 3})^2$. These contributions generate the on-shell
fourfold pole in Eq. (\ref{fourpoint}).
\end{itemize} 

\setcounter{equation}{0}
\section{\boldmath{$\pic\pim \rightarrow \pin\pin$} scattering amplitude}
\label{spipi}
As a check, we calculated the $\pic\pim \rightarrow \pin\pin$ amplitude to
$\order{(p^4, e^2p^2)}$ from (\ref{Zt}), (\ref{Zu2}) and (\ref{Zu3}). We
compared the thus obtained expression with the result of Knecht and Urech
\cite{Knecht&Urech}, who used a photon mass as an infrared regulator. Upon
identifying the infrared pole term in dimensional regularization with 
\begin{equation}
  \lIR = -\frac{1}{32\pi^2}\big(1+{\rm ln}\frac{m_\gamma^2}{\mu^2}\big),
\end{equation}
the two results are in agreement.
\subsection{Infrared singularities} 
The on--shell matrix element contains infrared singularities generated by
virtual photon contributions. Since we used dimensional regularization to
treat the infrared pole terms, the amplitude is well defined in $d\neq 4$
only. The essential loop function, which develops an infrared singularity as
$d\rightarrow 4$, is the scalar triangle diagram $G_d(s)$ with one photon and
two charged on--shell pion propagators. A discussion of the analytic
properties of $G_d(s)$, as well as an explicit representation of this function, is given in appendix \ref{triangle}.

Expanding the amplitude around threshold $s =4\Mpic^2$ leads to
\begin{equation}
  {\rm Re}A^{+-00} =
  -\frac{e^2}{16F^2}\frac{\Mpic}{|\vec{q}|}\left(3\Mpic^2+\Delta_\pi\right)+{\rm Re}A^{+-00}_{\rm thr}+\order{(|\vec{q}|)},
\end{equation}
in the Condon-Shortley phase conventions. Here, $\vec{q}$ denotes the relative 3-momentum of the charged pions and the Coulomb pole stems from the infrared region of the scalar triangle
diagram (\ref{Gdcoulomb}). The quantity ${\rm Re}A^{+-00}_{\rm thr}$ is free of infrared singularities.
\subsection{Amplitude at threshold}
We may now expand ${\rm Re}A^{+-00}_{\rm thr}$ in powers of the isospin breaking parameter $\delta$,
\begin{equation}
  \alpha \doteq \frac{e^2}{4\pi} = \order{(\delta)}, \quad (\Deltam)^2 = \order{(\delta)},
\end{equation}
according to
\begin{equation}
-\frac{3}{32\pi}{\rm Re}A^{+-00}_{\rm thr} = a_0-a_2 + h_1(\Deltam)^2 +
h_2\alpha +\order{(\delta^{3/2})}.
\end{equation}
The combination $a_0-a_2$ denotes the S-wave scattering length difference in the isospin symmetry limit $m_u = m_d$, $\alpha=0$ \cite{SU2},
\begin{equation}
  a_0-a_2 =
  \frac{9\Mpic^2}{32\pi
  F^2}\left[1+\frac{\Mpic^2}{288\pi^2F^2}\left(33+8\bar{l}_1+16\bar{l}_2-3\bar{l}_3\right)\right]+\order{(\hat{m}^3)}.
\end{equation}
The coefficients $h_1$ and $h_2$ of the isospin breaking contributions are
given by \cite{pipiatom, numpipiatom},  
\begin{eqnarray}
  h_1 &=& \order{(\hat{m})},\nn\\
  h_2 &=& \frac{3\Delta_\pi^{e.m.}}{32\pi\alpha
  F^2}\left[1+\frac{\Mpic^2}{12\pi^2F^2}\left(\frac{23}{8}+\bar{l}_1+\frac{3}{4}\bar{l}_3\right)\right]\nn\\
&& +\frac{3\Mpic^2}{256\pi^2F^2}p(k_i)+\order{(\hat{m}^2)},
\end{eqnarray}
where $\Delta_\pi^{e.m.}$ denotes the physical mass difference at $m_u = m_d$,
\begin{equation}
  \Delta_\pi^{e.m.} = \Delta_\pi \big|_{m_u = m_d},
\end{equation}
and $p(k_i)$ stands for the following combination of low--energy constants
\begin{equation}
  p(k_i) = -30+9\bar{k}_1+6\bar{k}_3+2\bar{k}_6+\bar{k}_8+\frac{4Z}{3}\left(\bar{k}_1+2\bar{k}_2+6\bar{k}_4+12\bar{k}_6-6\bar{k}_8\right).
\end{equation}

\setcounter{equation}{0}
\section{Summary and Outlook}
In the present paper we constructed an explicit representation of the
generating functional for ${\rm SU(2)_R} \times {\rm SU(2)_L} \times {\rm
  U(1)_V}$ with virtual photons at first non-leading order in the low--energy
expansion [$\order{(p^4, e^2p^2)}$]. The explicit form includes graphs, where up to two pions and at most one photon are running in the loop.

It has been shown in \cite{Knecht&Urech} that the ultraviolet divergences of
the one--loop functional are absorbed by the corresponding counterterms of
$\order{(p^4,e^2p^2)}$. For non-exceptional 4-momenta the generating
functional is thus finite as $d\rightarrow4$. Nevertheless, if we want to
extract on--shell matrix elements, we encounter in general infrared singularities, generated by soft photon contributions. We use dimensional
regularization to treat these infrared pole terms and therefore specify the explicit form of the generating functional in $d \neq 4$ dimensions.

As a check of the one--loop functional, we evaluated the $\pic\pim \rightarrow
\pin\pin$ scattering amplitude at order $p^4, e^2p^2$. The thus obtained result
agrees with the one in Ref. \cite{Knecht&Urech}. 

Furthermore, we specify the electromagnetic corrections to the decay constants
$F_\pi$ and $G_\pi$, which measure the one-particle matrix elements of the isovector axial and pseudoscalar currents. 

The extension to ${\rm SU(3)_R} \times {\rm SU(3)_L}$ is in progress
\cite{SU3&photons}. It will provide a comparison with the $\pi-K$ amplitudes
already available in the literature \cite{piK}.
\section*{Acknowledgments}
I am grateful to J. Gasser for many illuminating discussions as well as for
helpful comments on the manuscript. Further, I thank R. Kaiser for useful remarks. This work
was supported in part by the Swiss National Science Foundation, and by
RTN, BBW-Contract N0.~01.0357 and EC-Contract HPRN-CT-2002-00311
(EURIDICE).
\appendix

\renewcommand{\thesection}{\Alph{section}}
\renewcommand{\theequation}{\Alph{section}.\arabic{equation}}
\setcounter{section}{0}
\section{Kernels}
\label{Kernels}
Here, we explain the notation used in subsection \ref{sexplZ}. We consider the
Fourier transforms ($s = p^2$)
\begin{eqnarray}
  K_\mu(p, M_i^2, M_k^2) &=& \int d^d z e^{i p z}K_\mu(z, M_i^2, M_k^2),\nn\\
M_{\mu\nu}^r(p, M_i^2, M_k^2) &=& \int d^d z e^{i p z}M_{\mu\nu}^r(z, M_i^2, M_k^2).
\end{eqnarray}
The kernel $K_\mu$ remains ultraviolet finite as $d\rightarrow 4$,
\begin{equation}
  K_\mu(p, M_i^2, M_k^2) = 
  \frac{i p_\mu}{2}\frac{M_i^2-M_k^2}{s}\bar{J}(s,M_i^2, M_k^2),
\end{equation}
where $\bar{J}(s)=J(s)-J(0)$ and 
\begin{equation}
  J(s, M_i^2, M_k^2)  = \frac{1}{i}\int \frac{d^d q}{(2\pi)^d} \frac{1}{(M_i^2-q^2)(M_k^2-(p-q)^2)}.
\end{equation}
The renormalized kernel $M^r_{\mu\nu}$ reads:
\begin{equation}
  M^r_{\mu\nu}(p, M_i^2, M_k^2) = -(p_\mu p_\nu-g_{\mu\nu} s)M^r(s, M_i^2,
  M_k^2)-g_{\mu\nu}L(s, M_i^2, M_k^2),
\end{equation}
with
\begin{eqnarray}
  M^r(s, M_i^2, M_k^2) &=&
  \frac{1}{4(d-1)s}\Big\{\big[s-2(M_i^2+M_k^2)\big]\bar{J}(s, M_i^2,
  M_k^2)\nn\\
&& +\frac{d}{s}(M_i^2-M_k^2)^2\bar{{\bar{J}}}(s, M_i^2, M_k^2)\Big\}-\frac{1}{6}k(M_i^2, M_k^2)+\frac{1}{288\pi^2},\nn\\
L(s, M_i^2, M_k^2) &=& \frac{1}{4s}(M_i^2-M_k^2)^2\bar{J}(s, M_i^2, M_k^2),
\end{eqnarray}
and
\begin{eqnarray}
  \bar{\bar{J}}(s) &=& \bar{J}(s)-s\bar{J}^\prime(0),\nn\\
  k(M_i^2, M_k^2) &=& \frac{1}{32\pi^2(M_i^2-M_k^2)}\Big[M_i^2{\rm ln}\frac{M_i^2}{\mu^2}- M_k^2{\rm ln}\frac{M_k^2}{\mu^2}\Big].
\end{eqnarray}
We list the ultraviolet finite kernels $K_\mu$ and $M^r_{\mu\nu}$ in $d \neq
4$ dimensions, because we encounter an infrared singularity, if we expand the
charged pion self-energy around its physical mass. This pole term,
which comes from $\bar{J}^\prime (\Mpic^2, \Mpic^2,0)$, leads to an infrared
divergent scaling factor $Z_+$ in Eq. (\ref{Zfactor}).  

Furthermore, we introduce the Fourier transform
\begin{equation}
  T(p_1,p_2) = \int d^d u \, d^d v \, e^{i p_1
  u}e^{i p_2 v} T(u,v),
\end{equation}
and similarly for the other kernels $T_{i\mu}, T_{i\mu\nu}$ and
$T_{\mu\nu}^\mu$ in Eq. (\ref{Zu3}). Then we have
\begin{eqnarray}T(p_1, p_2) &=& I, \nn\\
T_{1\mu}(p_1, p_2) &=& i p_{1\mu} I-2i I_\mu, \nn\\
T_{2\mu}(p_1, p_2) &=& i p_{2\mu} I-2i I_\mu, \nn\\
T_{3\mu}(p_1, p_2) &=& i (p_1+p_2)_\mu I-2i I_\mu, \nn\\
T_{1\mu\nu}(p_1, p_2) &=& 4 I_{\mu\nu}-2(p_1+p_2)_\mu I_\nu-2p_{1\nu}
I_\mu +(p_1+p_2)_\mu p_{1\nu} I, \nn\\
T_{2\mu\nu}(p_1, p_2) &=& 4 I_{\mu\nu}-2(p_1+p_2)_\nu I_\mu-2p_{2\mu}
I_\nu +p_{2\mu}(p_1+p_2)_\nu I, \nn\\
T_{3\mu\nu}(p_1, p_2) &=& 4 I_{\mu\nu}-2p_{1\mu} I_\nu-2p_{2\nu}
I_\mu +p_{1\mu} p_{2\nu} I, \nn\\
T_{\mu\nu}^\mu(p_1, p_2) &=& i\left\{-8I_{\mu\nu}^\mu+4(p_1+p_2)^\mu
  I_{\mu\nu}+4(p_1+p_2)_\nu I^\mu_\mu-2(p_1\cdot p_2) I_\nu\right. \nn\\
&& \left. -2(p_1+p_2)_\mu(p_1+p_2)_\nu
  I^\mu + (p_1 \cdot p_2)(p_1+p_2)_\nu I\right\},
\end{eqnarray}
with
\begin{equation}
  \{I,I_\mu,I_{\mu\nu},I_{\mu\nu\rho}\} = \frac{1}{i}\int \frac{d^d
  k}{(2\pi)^d}\frac{\{1, k_\mu, k_\mu k_\nu, k_\mu k_\nu k_\rho\}}{(M_i^2-(p_1-k)^2)(M_k^2-(p_2-k)^2)(-k^2)}.
\end{equation}
The integrals $I$ and $I_\mu$ are ultraviolet finite, but $I$
develops an infrared singularity for on--shell 4-momenta (see appendix
\ref{triangle}). The functions $I_{\mu\nu}$ and $I_{\mu\nu\rho}$ have
ultraviolet poles at $d=4$: 
\begin{eqnarray}
  T_{i \mu\nu}(p_1, p_2)&=& T_{i \mu\nu}^r(p_1, p_2)+2g_{\mu\nu}\lambda\quad
  i = 1, \dots, 3, \nn\\
  T_{\mu\nu}^\mu(p_1, p_2)&=& T_{\mu\nu}^{\mu r}(p_1, p_2)+
  \tfrac{2}{3}i\lambda (d-1)(p_1+p_2)_\nu. 
\end{eqnarray}
For non-exceptional 4-momenta the renormalized quantities $T_{i\mu\nu}^r$ and
$T_{\mu\nu}^{\mu r}$ remain finite as $d\rightarrow 4$.

\setcounter{equation}{0}
\section{Scalar vertex function}
\label{triangle}
The scalar vertex function $G_d(s)$ is given by 
\begin{equation}
  G_d(s) = \frac{1}{i}\Int \frac{d^d
  k}{(2\pi)^d}\frac{1}{\Mpic^2-(p_1-k)^2}\frac{1}{\Mpic^2-(p_2+k)^2}\frac{1}{-k^2},
\label{Gd}
\end{equation} 
where $p_1$, $p_2$ denote the on--shell 4-momenta of the incoming $\pic$,
$\pim$ mesons and $s= (p_1+p_2)^2$. The vertex function is ultraviolet finite,
but develops an infrared singularity at $d = 4$. (At $p_i^2 \neq \Mpic^2$, the
singularity would be absent).

\begin{figure}[h t]
\begin{center}
\thispagestyle{empty}
\setlength{\unitlength}{1mm} 
\begin{picture}(40,40)
\put(0,0){\vector(1,0){40}}
\put(0,37){\makebox(0,0){$d$}}
\put(0,0){\vector(0,1){35}}
\put(42,0){\makebox(0,0){$s$}}
\put(0,20){\line
(1,0){40}}
\put(-5,20){\makebox(0,0){$d=4$}}
\put(20,0){\line(0,1){35}}
\put(23.5,-3){\makebox(0,0){$s=4M_{\pi_+}^2$}}
\put(0,0){\line(1,1){19.8}}
\put(0,5){\line(1,1){14.8}}
\put(0,10){\line(1,1){9.8}}
\put(0,15){\line(1,1){4.8}}
\put(5,0){\line(1,1){19.8}}
\put(10,0){\line(1,1){19.8}}
\put(15,0){\line(1,1){19.8}}
\put(20,0){\line(1,1){19.8}}
\put(25,0){\line(1,1){14.8}}
\put(30,0){\line(1,1){9.8}}
\put(35,0){\line(1,1){4.8}}
\thicklines 
\put(30,30){\vector(-1,0){7}}
\put(23,30){\line(-1,0){3}}
\put(30,30){\circle*{1.2}}
\put(30,30){\vector(0,-1){7}}
\put(30,23){\line(0,-1){3}}
\put(30,20){\vector(-1,0){10}}
\put(20,30){\vector(0,-1){10}}
\put(17.5,25){\makebox(0,0){$b$}}
\put(32.5,25){\makebox(0,0){$a$}}
\end{picture}
\caption{\label{fig: Gd}Two different paths to approach to threshold. The
  shaded region is excluded since the integral $G_d$ does converge for $d>4$ only.}
\end{center}
\end{figure}
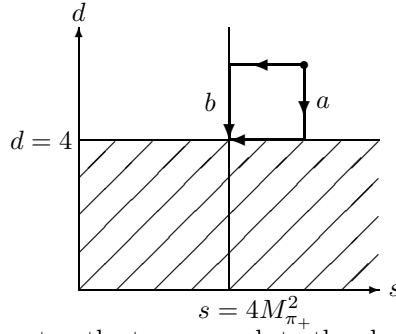
For $s\geq 4\Mpic^2$, the function has a cut along the positive real axis. Expanding the vertex function in $d-4$ for arbitrary $s\geq 4\Mpic^2$ yields
\begin{eqnarray}
 G_d(s) &=&
 \frac{1}{32\pi^2 s\sigma}\bigg\{2\pi^2+4 {\rm Li}\frac{1+\sigma}{1-\sigma}+{\left({\rm ln}\frac{1+\sigma}{1-\sigma}\right)}^2\nn\\
&&-2{\rm ln}\frac{1+\sigma}{1-\sigma}\left(1+32\pi^2 \lIR+{\rm
 ln}\frac{\Mpic^2}{\mu^2}\right)\bigg\}\nn\\
&&+\frac{i}{16\pi s \sigma}\left(1+32\pi^2\lIR+{\rm
 ln}\frac{s\sigma^2}{\mu^2}\right)+\order{(d-4)},
\label{Gds}
\end{eqnarray}
with 
\begin{equation}
  \sigma = \sqrt{1-\frac{4\Mpic^2}{s}},
\end{equation}
and 
\begin{equation}
  {\rm Li}(z) \doteq {\rm Li_2}(1-z) = \int_1^z du \frac{{\rm ln}u}{1-u}.
\end{equation}
If we expand (\ref{Gds}) near threshold $s = 4\Mpic^2 + 4\vec{q}^2$, which corresponds to path $a$ in Fig. \ref{fig: Gd}, we find
\begin{eqnarray}
  G_d(s)
  &=&\frac{1}{64\Mpic^2}\bigg\{\frac{\Mpic}{|\vec{q}|}-\frac{2}{\pi^2}\left(3+32\pi^2\lIR+{\rm ln}\frac{\Mpic^2}{\mu^2}\right)-\frac{|\vec{q}|}{2\Mpic}\nn\\
&&+
  \frac{i}{\pi}\left(\frac{\Mpic}{|\vec{q}|}-\frac{|\vec{q}|}{2\Mpic}\right)\left(1+32\pi^2\lIR+{\rm ln}\frac{4\vec{q}^2}{\mu^2}\right)\bigg\}\nn\\
&&+\order{(d-4)}+\order{(\vec{q}^2)},
\label{Gdcoulomb}
\end{eqnarray}
where $\vec{q}$ denotes the relative momentum of the incoming pions in the CM frame. The singular contributions $\sim 1/|\vec{q}|$ are generated by small
values of the variable of integration $k$ in (\ref{Gd}). 

On the contrary, if we first evaluate the vertex function at threshold in $d \neq 4$ dimensions
\begin{equation}
  G_d(4\Mpic^2)= \frac{C_d}{\Mpic^2}\frac{1}{(d-4)(d-5)}\Mpic^{d-4}, \quad 
  C_d = \frac{1}{(4\pi)^{\frac{d}{2}}}\Gamma(3-\frac{d}{2}),
\end{equation}
and then expand in $d-4$, i.e we choose path $b$ in Fig. \ref{fig: Gd}
\begin{equation}
   G_d(4\Mpic^2) = -\frac{1}{\Mpic^2}\left[\lIR+\frac{1}{32\pi^2}\left(3+{\rm ln}\frac{\Mpic^2}{\mu^2}\right)\right]+\order{(d-4)},
\end{equation}  
we are left with the constant term in Eq. (\ref{Gdcoulomb}).

\setcounter{equation}{0}
\section{Coupling constants \boldmath{$F_\pi$} and \boldmath{$G_\pi$} }
\label{decayconstants}
In presence of electromagnetic interactions the coupling constants of the
isovector axial and pseudoscalar currents to the pion are given by\footnote{The coefficient of $\sum c_i \bar{k}_i$ in $F_{\pin}$ differs by a
  factor $1/2$ from the corresponding expression in Ref. \cite{Knecht&Urech}.},
\begin{eqnarray}
  F_{\pic} &=& F_\pi+F\bigg\{ 2e^2 \big[\lIR+\frac{1}{32\pi^2}\big({\rm
  ln}\frac{\Mpic^2}{\mu^2}-2\big)\big]-\frac{e^2Z}{8\pi^2}\bar{l}_4\nn\\
 &&  -\frac{e^2}{64\pi^2}\left[
  \left(3+\tfrac{4Z}{9}\right)\bar{k}_1 -\tfrac{40Z}{9}\bar{k}_2 -
  \bar{k}_9\right]\nn\\
&& +\frac{\Mpic^2-2Z e^2 F^2}{32\pi^2 F^2}{\rm
  ln}\frac{\Mpic^2}{\Mpin^2}+\order{(e^2 m_q)}\bigg\}, \nn\\
 F_{\pin} &=& F_\pi-F\bigg\{ \frac{e^2
  Z}{8\pi^2}\bar{l}_4 +
  \frac{e^2}{64\pi^2}\left[\left(3+\tfrac{4Z}{9}\right)\bar{k}_1-\tfrac{40Z}{9}\bar{k}_2 \nn \right.\\
&& \left. -3 \bar{k}_3-4Z \bar{k}_4\right]+\order{(e^2 m_q)}\bigg\},
\end{eqnarray}
and
\begin{eqnarray}
G_{\pic} &=& G_\pi+2BF\bigg\{2e^2\big(\lIR+\frac{1}{32\pi^2}{\rm
  ln}\frac{\Mpic^2}{\mu^2}\big)+\frac{e^2Z}{16\pi^2}\bar{l}_3\nn\\
&&+\frac{e^2}{64\pi^2}\left[\left(3+\tfrac{4Z}{9}\right)\bar{k}_1-\tfrac{40Z}{9}\bar{k}_2-\tfrac{2}{9}\left(5+4Z\right)\bar{k}_5\right.
  \nn\\
&&\left.+\tfrac{10}{9}\left(1+8Z\right)\bar{k}_6+\left(1-8Z\right)\bar{k}_8\right]
  +\frac{4e^2}{9}k_7+\order{(e^2 m_q)}\bigg\}, \nn\\
G_{\pin} &=& G_\pi+2BF\bigg\{\frac{e^2Z}{16\pi^2}\left(\bar{l}_3-2\bar{l}_4\right)+\frac{e^2}{64\pi^2}\left[\left(3+\tfrac{4Z}{9}\right)\bar{k}_1-\tfrac{40Z}{9}\bar{k}_2\right.\nn\\
&&\left.-3\bar{k}_3-4Z\bar{k}_4-\tfrac{2}{9}\left(5+4Z\right)\bar{k}_5
+\tfrac{10}{9}\left(1+8Z\right)\bar{k}_6\right]+\frac{4e^2}{9}k_7\nn\\
&&+\frac{\Mpic^2-2e^2F^2Z}{32\pi^2F^2}{\rm ln}\frac{\Mpic^2}{\Mpin^2}
  +\order{(e^2 m_q)}\bigg\}.
\end{eqnarray}
The quantities $F_\pi$ and $G_\pi$ denote the coupling constants at $\alpha =
0$ and $m_u = m_d$ \cite{SU2},
\begin{eqnarray}
F_\pi &=& F\big[1+\frac{\Mpic^2}{16\pi^2 F^2}\bar{l}_4+\order{(m_q^2)}\big],\nn\\
G_\pi &=&
  2BF\big[1-\frac{\Mpic^2}{32\pi^2F^2}\left(\bar{l}_3-2\bar{l}_4\right)+\order{(m_q^2)}\big].
\end{eqnarray}
Due to the infrared pole term, $F_{\pic}$ and $G_{\pic}$ are well defined in
$d \neq 4$ dimensions only. 

The coupling constant of the isoscalar $P^0$ to the pion is of $\order{((\Deltam), e^2)}$
\begin{equation}
  \tilde{G}_\pi = \frac{4B^2 (\Deltam)}{F}l_7 + \frac{8B F e^2}{3}k_7.
\end{equation} 

\end{document}